\documentstyle[twocolumn,seceq,psfig19]{jpsj}
\def\sf{\rm}
\renewcommand\figureheight[1]{\vspace{24pt}\mbox{\rule{0cm}{#1}}}
\newcommand{\be}{\begin{equation}} 
\newcommand{\ee}{\end{equation}}
\newcommand{\bd}{\begin{displaymath}}
\newcommand{\ed}{\end{displaymath}}
\newcommand{\ba}{\begin{eqnarray}}
\newcommand{\ea}{\end{eqnarray}}

\def\gsim{\lower3pt\hbox{$\stackrel{>}{\scriptstyle \sim}$}} %greater, sim.
\def\lsim{\lower3pt\hbox{$\stackrel{<}{\scriptstyle \sim}$}} %less, sim.
\def \theequation{\thesection.\arabic{equation}}

\def\x{\mib{x}}
\def\r{\mib{r}}

\def\e{{\rm e}}
\def\t{\theta}
\def\l{\phi}

\def\no{\nonumber}
\def\av#1{{\left\langle#1\right\rangle}}

\def\runtitle{Conditional Averages and PDF's in the Passive Scalar Field}
\def\runauthor{Naoya {\sc Takahashi}, Tsutomu {\sc Kambe}, Tohru {\sc Nakano}, Toshiyuki {\sc Gotoh}, 
  and Kiyoshi {\sc Yamamoto}}
\title
{
Conditional Averages and Probability Density Functions\\ in the Passive Scalar Field
}

\author
{
Naoya {\sc Takahashi},$^{1}$\footnote{E-mail address:takahasi@swift.phys.s.u-tokyo.ac.jp}
Tsutomu {\sc Kambe},
Tohru {\sc Nakano},\\ 
Toshiyuki {\sc Gotoh},
and Kiyoshi {\sc Yamamoto}
}

\inst
{
$^1$ Department of Physics, University of Tokyo, Tokyo 113,\\
$^2$ Department of Physics, Chuo University, Tokyo 112,\\
$^3$ Department of System Engineering, Nagoya Institute of Technology, Nagoya 466,\\
$^4$ National Aerospace Laboratory, Chofu, Tokyo 182
}

\recdate
{
\hspace{5cm}
}

\abst
{
Two conditional averages for the increment $\Delta(\x,\x+\r) \equiv
T(\x)-T(\x+\r)$ of the scalar field are estimated for DNS data:
$H(\Delta)=\av{\nabla_{\x}^2 \Delta(\x,\x+\r)|\Delta(\x,\x+\r)}$ and
$G(\Delta)=\av{|\nabla_{\x} \Delta(\x,\x+\r)|^2| \Delta(\x,\x+\r)}$. 
The probability density function $P(\Delta)$ for $\Delta(\x,\x+\r)$ is also calculated.  If
$\Delta$, $P(\Delta)$, $H(\Delta)$, and $G(\Delta)$ are expressed as $\t$, $P(\t)$,
$h(\t)$, and $g(\t)$ in a dimensionless form,
the data analysis has revealed interesting simple relationships between them:  
(1) $g(\t)P(\t) \sim \exp(-\int_0^{\t} du h(u)/g(u))$ derived by Ching and Kraichnan
was numerically confirmed perfectly,  
(2) $g(\t)P(\t)$ pass through an almost common point for any scale ranging
from the dissipative to the inertial, to the energy containing scale,
(3) by using fitting parameters $A$ and $\beta$ as a function of $r$,
$h(\t)/g(\t)$ is satisfactorily fitted by $(A/\beta)\tanh(\beta \t)$,
and then $g(\t)P(\t) \sim   (\cosh \beta \t)^{-A/\beta^2}$.
These relations may be helpful to construct a turbulence model. 
}

\kword
{
conditional averages, pdf, passive scalar field, anomalous scaling
}

\begin{document}
\sloppy
\maketitle

\section{Introduction}

Since Kraichnan's work,~\cite{rf:k1} the anomalous scaling
of the passive scalar field has become a 
prototype of  a study of the intermittency problem of turbulence. ~\cite{rf:lpf,rf:gk,rf:cfkl,rf:ss}
The scaling is evaluated by the scaling exponents of the $n$th order structure functions $S_{n}(r)$
of the scalar field $T(\x)$, which are expressed as
\be
    S_n(r)=\av{\Delta(\x,\x+\r)^n},
\ee
where $\Delta(\x,\x+\r)$ is the increment of the scalar field denoted by
\be
    \Delta(\x,\x')=T(\x)-T(\x').
\ee

If the scalar filed is advected by the $\delta$-correlated Gaussian velocity field, $S_{n}(r)$ was 
shown to be governed by the following simple equation in the steady state~\cite{rf:k1,rf:k2}: 
\be
    D_{n}(r)=J_{n}(r), \label{dj}
\ee
where $D_{n}(r)$ is the inertial term expressed by
\be
   D_{n}(r)=-r^{1-d} \frac{\partial}{\partial r} \left( a(r) r^{d-1} \frac{\partial}{\partial r} 
     S_{n}(r) \right), \label{d}
\ee
where $a(r)$ is the eddy diffusivity, and $d$, the dimension of the system.  $J_{n}(r)$ is the
dissipative term defined by
\be
  J_{n}(r)=n \kappa \av{\Delta(\x,\x')^{n-1}\Lambda(\x,\x')},  \label{j1}
\ee
where 
\be
   \Lambda(\x,\x')=(\nabla_{\x}^2+\nabla_{\x'}^2)\Delta(\x,\x'). \label{Lambda}
\ee
In eq.(\ref{j1}) $\kappa$ is the molecular diffusion coefficient and $\x'$ is put equal to $\x+\r$.  

In order to close the equation (\ref{dj}), one has to make a certain approximation on $J_{n}(r)$. 
The ansatz, which was employed by Kraichnan,~\cite{rf:k1} is that the conditional average of the
field $\Lambda(\x,\x')$ with  a value of $\Delta(\x,\x')$ being fixed is proportional to
$\Delta(\x,\x')$ itself, i.e.
\ba
     H(\Delta(\x,\x'))&=&\av{\Lambda(\x,\x')|\Delta(\x,\x')} \no\\
                      &=&f_1(r)\Delta(\x,\x'), \label{linear}
\ea
where $f_1(r)=J_2(r)/(2\kappa S_2(r))$.  Under this ansatz $J_{n}(r)=n J_2(r)S_{n}(r)/(2S_2(r))$,  
which enables us to close the equation (\ref{dj}).  Although this ansatz was examined in simulation
and experiment,~\cite{rf:kyc,rf:fglp,rf:clp,rf:clpp} the definite confirmation has not been reached
yet.  Hence, it is interesting to examine to what extent of $\Delta$ the linear relation
(\ref{linear}) is satisfied.

Another conditional average is conceivable.  Making a partial integration in $J_{n}(r)$ with respect
to $\x$,  we  have
\be
   J_{n}(r)=-n(n-1) \kappa \av{\Delta(\x,\x+\r)^{n-2} \Sigma(\x,\x+\r)}, \label{j2}
\ee
where
\be
    \Sigma(\x,\x+\r)=|\nabla_{\x}(T(\x)-T(\x+\r))|^2. \label{sigma}
\ee
Now we introduce another conditional average
of $\Sigma(\x,\x+\r)$ with a fixed value of $\Delta(\x,\x+\r)$ as
\be
    G(\Delta(\x,\x+\r))=\av{\Sigma(\x,\x+\r)|\Delta(\x,\x+\r)} \label{G}.
\ee

Although the linear ansatz has not been proved yet, it might have an origin in the
dynamics of  $\Delta(\x,\x+\r)$.  On the other hand, the conditional average (\ref{G}) might be 
closely related to the conservation of scalar dispersion, since $\Sigma(\x,\x+\r)$ represents the scalar diffusion
rate.  Which average is more appropriate will be judged through the examination of the  result.  
This is the first aim of the present paper.   

The second motivation of examining those averages is the recent argument by Ching and
Kraichnan,~\cite{rf:ck} which  enables us to combine $H$ and $G$ to yield the probability
distribution function (pdf) for $\Delta(\x,\x')$.  They showed that the pdf is related to the $H/G$ in
the homogeneous situation.  We will confirm the relationship perfectly later in this paper.  In this
sense the conditional averages $H$ and $G$ are expected to shed light on the pdf of the increment of
the scalar field and the issue of intermittency of turbulence.

Although the problem of the anomalous scaling should be ideally investigated in the randomly 
generated velocity fields, the generation of such fields is very difficult.  Thus, the above
conditional averages are examined in the usual Navier-Stokes turbulence in the present paper.

Finally we emphasize that the present paper is not directly 
devoted to the study of the recent issue of the anomalous scaling. 
One of the important features in the study is the introduction of 
the conditional average of spatial derivatives of the scalar field  
with a given value of its difference.  
Combination of the conditional average with the equation of motion 
and/or some  kinematical constraints yields various check points to examine 
the pdf for the scalar difference, 
which would lead us to a more precise knowledge of the pdf. 
The point of view is not lmitted only to the case for the scalar 
field advected by the velocity field white in time. 
In this sense, we examine the statistics of the scalar field
advected by the Navier-Stokes turbulence~\cite{rf:cc} from 
the different point of angle with the help of the
methods developed in the area of the anomalous scaling.

The present paper is organized as follows.  In the following section the algorithm of the simulation
as well as the statistical property of the velocity field is given.  In \S3 the
pdf's of the fields $\Delta$ and $\Lambda$ are presented with emphasis on their scale and 
Reynolds number dependence.  To what extent of $\Delta$ the linear hypothesis of 
$H$ is valid is examined in \S4 and it is shown that it does not hold for large values of
$\Delta$, which correspond to the structure functions of orders larger than three.  The
deviation from the linearity is also discussed.  In \S5 the dependence of $G$ on $\Delta$ is
investigated; it is described by a simple fitting formula, which is valid up to structure functions
of large orders.  Section 6 is devoted to the pdf of $\Delta$ derived from the combination of $H$
and $G$.  The numerically obtained pdf agrees with the observed one perfectly.  The remarkable
findings about $H/G$ and $P(\Delta) G$ are given and the results are analytically
expressed using parameters.  In \S7 the useful findings are summarized to help to construct a
turbulence model.

\section{Simulation}

The simulation was done with resolutions of $256^3$ (Run 1 and 2) and $512^3$ (Run 3) mesh points for
several values of  molecular viscosity using the NAL Numerical Wind Tunnel.~\cite{rf:y}  The velocity
and scalar  field are simulated at the same time in $4\pi$ cyclic box by making use of the
Navier-Stokes equation and the advection equation. The Prandtl number is set equal to unity. The
turbulence decays from the same initial spectrum for velocity and scalar fields
\begin{equation}
        \frac{16}{3} \sqrt{ \frac{2}{\pi}} k^4 \exp\left(-2k^2\right).
\end{equation}
The data were processed at t=10, when the energy dissipation rate reaches the largest value through 
the simulation, but the scalar one has passed the largest value already.  At that time the Taylor
microscale Reynolds number $R_{\lambda}$ ranges from 89.5 for Run 1 to 120 for Run 3.  The inertial
region is identified from the scalar spectrum and the requirement of the relation $T_r^2u_r \sim r$,
where $T_r$ and $u_r$ are the scalar and velocity increment over $r$.  The related data are
summarized in Table 1.

\begin{table}
\caption{Statistical property of the present DNS.  Run 1 and 2 are carried on $256^3$ mesh points and 
Run 3 is on $512^3$ mesh points.  $L_p$ is the integral scale,
$\lambda$, the Taylor microscale, and $\eta$, the Kolmogorov microscale.  $k_{max}$ is the maximum
wavenumber after eliminating the aliasing error.}
\label{table:1}
\figureheight{4cm}
\begin{tabular}{@{\hspace{\tabcolsep}\extracolsep{\fill}}cccccc} \hline
 & $R_{\lambda}$ & $k_{max} \eta$ & $\lambda/\eta$ & $L_p/\eta$ & \mbox{inertial range} \\ \hline
 Run 1 & 89.5 & 1.27 & 18.6 & 63.4 & $7< r/\eta <31$ \\ \hline
 Run 2  & 120 & 0.716 & 21.5 & 68.4 & $12 < r/\eta <46 $ \\ \hline
 Run 3  & 120 & 1.42 & 21.8 & 103 & $14 < r/\eta <51$ \\ \hline
\end{tabular}
\end{table}

Since the resolution for Run 1 and 3 are sufficient, the data associated with Run 1 and 3 are 
mainly discussed.  The inertial region of Run 1 is located in 31 $> r/\eta >$7 for Run 1 and 51
$> r/\eta >$ 14 for Run 3, $\eta$ being the Kolmogorov dissipation scale. The dissipation spectrum is the largest at
about $r/\eta$ = 5 for both  runs, so that the scale $r/\eta$ = 5, which will appear often later, is in the dissipative
region.

Since Run 2 and Run 3 have the same Reynolds number with different resolution, the comparison of
the result of Run 2 with that of Run 3 is interesting to know the effect of the difference of the
resolution.

\section{The pdf's of the fields $\Delta_r$ and $\Lambda_r$}

\begin{figure}%--------------------------------------------------
%\figureheight{12cm}
%\epsfile{file=Fig1PDFreta20.eps,width=75mm}
\psfig{file=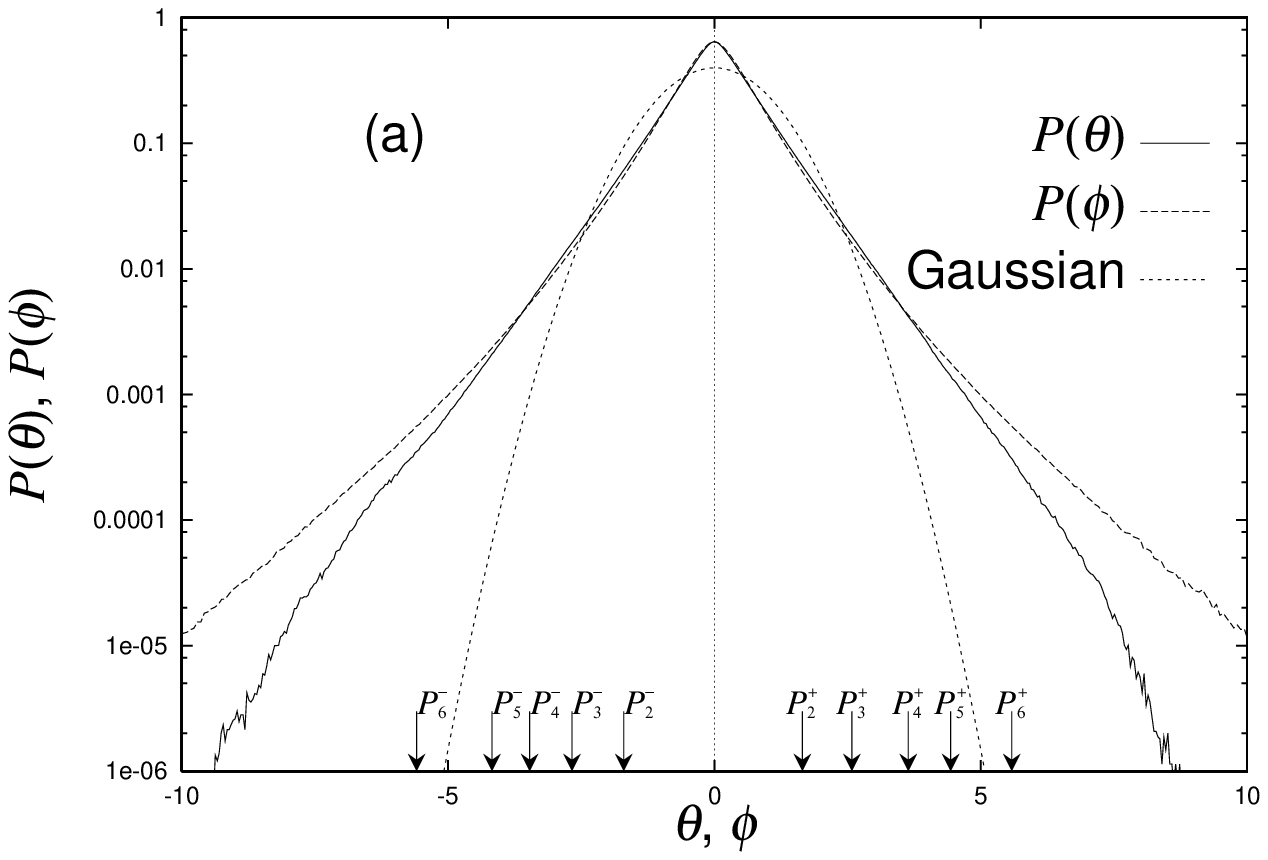,width=75mm}

%\epsfile{file=Fig2PDFreta05.eps,width=75mm}
\psfig{file=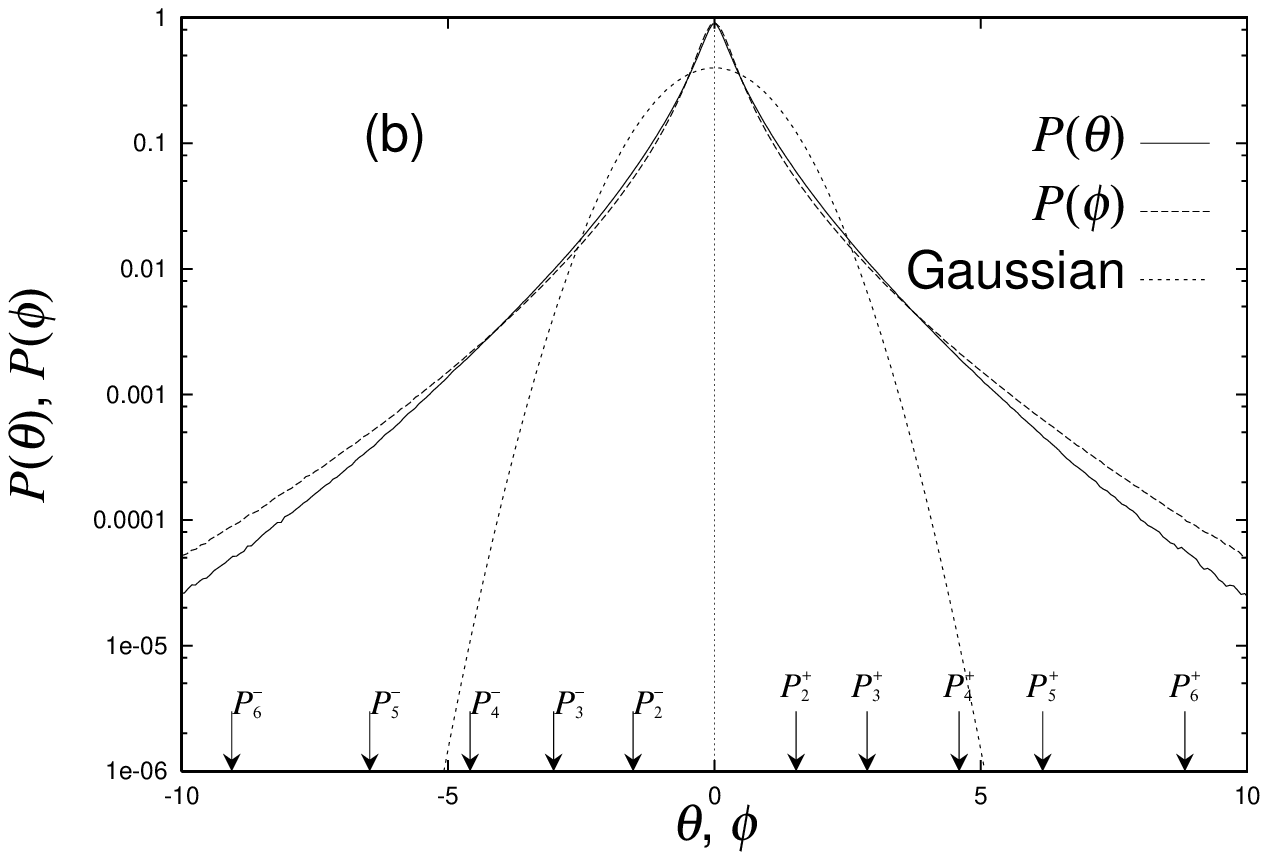,width=75mm}
\caption{The pdf's of $\theta$ and $\l$ for Run 3:  (a)
$r/\eta=20$ in the inertial region,  and (b) $r/\eta=5$ in the
dissipative region.  The vertical
quantities as well as the horizontal ones are normalized.  The solid line denotes $P(\t)$, while the
broken line does $P(\l)$.  The dotted line stands for the Gaussian distribution.  The arrow with
$P_n^+ $ signifies a peak position where $P(\t) \t^n$ is the largest for $\t>0$, while $P_n^-$
is a peak position where $P(\t) |\t|^n$ is the largest for $\t<0$. } 
\label{fig:1}
\end{figure}

In this section we will be concerned with the pdf's for the scalar field increment 
$\Delta_r(\x)=\Delta(\x,\x+\r)$ over two points separated by $r$ and the Laplacian field increment 
$\Lambda_r(\x)=\Lambda(\x,\x+\r)$.  
For simplicity we introduce the normalized dimensionless quantities such as 
\ba
   & &  \t_r(\x)=\frac{1}{\sqrt{S_2(r)}} \Delta_r(\x), \label{theta}\\
   & &  \l_r(\x)=\frac{2\kappa \sqrt{S_2(r)}}{J_2(r)}\Lambda_r(\x). \label{lambda}
\ea
The pdf's $P(\t_r)$ and $P(\l_r)$ of the fields $\t_r$ and $\l_r$ are obtained for various values of 
$r$ ranging from the 
dissipative scale to the upper inertial one.  From now on, we suppress the subscript $r$ without 
otherwise stated.  The pdf's for Run 3 are depicted in Fig.1, although the same ones could be  drawn
for Run 1; $r/\eta=20$ in the inertial region in Fig.\ref{fig:1}(a), and $r/\eta=5$ in the
dissipative region in Fig.\ref{fig:1}(b).  The vertical quantities as well as the horizontal ones
are normalized.  The solid line denotes $P(\t)$, while the broken line does $P(\l)$.  The dotted line
stands for the Gaussian distribution.  In order to emphasize which part of the fluctuations
contributes to the $n$th order structure functions, the arrows with $P_n^+$ and $P_n^-$ are inserted:
$P_n^+$ signifies a position where $P(\t) \t^n$ is the largest for $\t>0$ and $P_n^-$ does a
position where $P(\t) |\t|^n$ is the largest for $\t<0$.  

The scale-dependence of pdf's of $\t$ and $\l$ is different.  The frequency of large amplitudes of
$\t$ becomes less as $r$ increases, which implies that the degree of the intermittency measured from
$P(\t)$ is weaken with the increase of the scale $r$.  On the other hand, the pdf of $\l$ does not
vary so much as compared with that of $\t$  even when the scale changes.  This difference is
consistent with the fact that $\t_r$ represents the fluctuations of scale $r$, while $\l_r$ does the
ones of dissipative scale.

\begin{figure}
%\figureheight{12cm}
%\epsfile{file=Fig5.eps,width=75mm}
%
%\epsfile{file=Fig6.eps,width=75mm}
\psfig{file=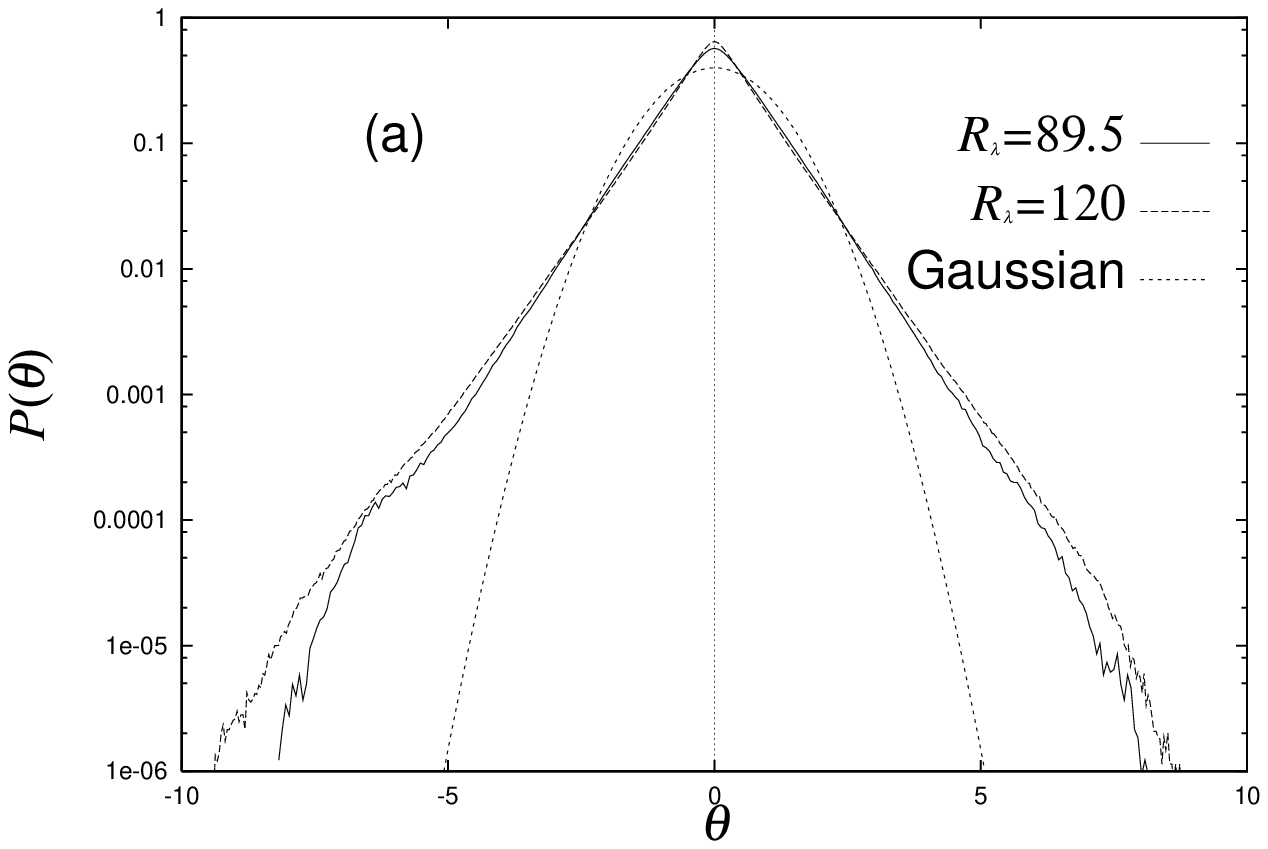,width=75mm}

\psfig{file=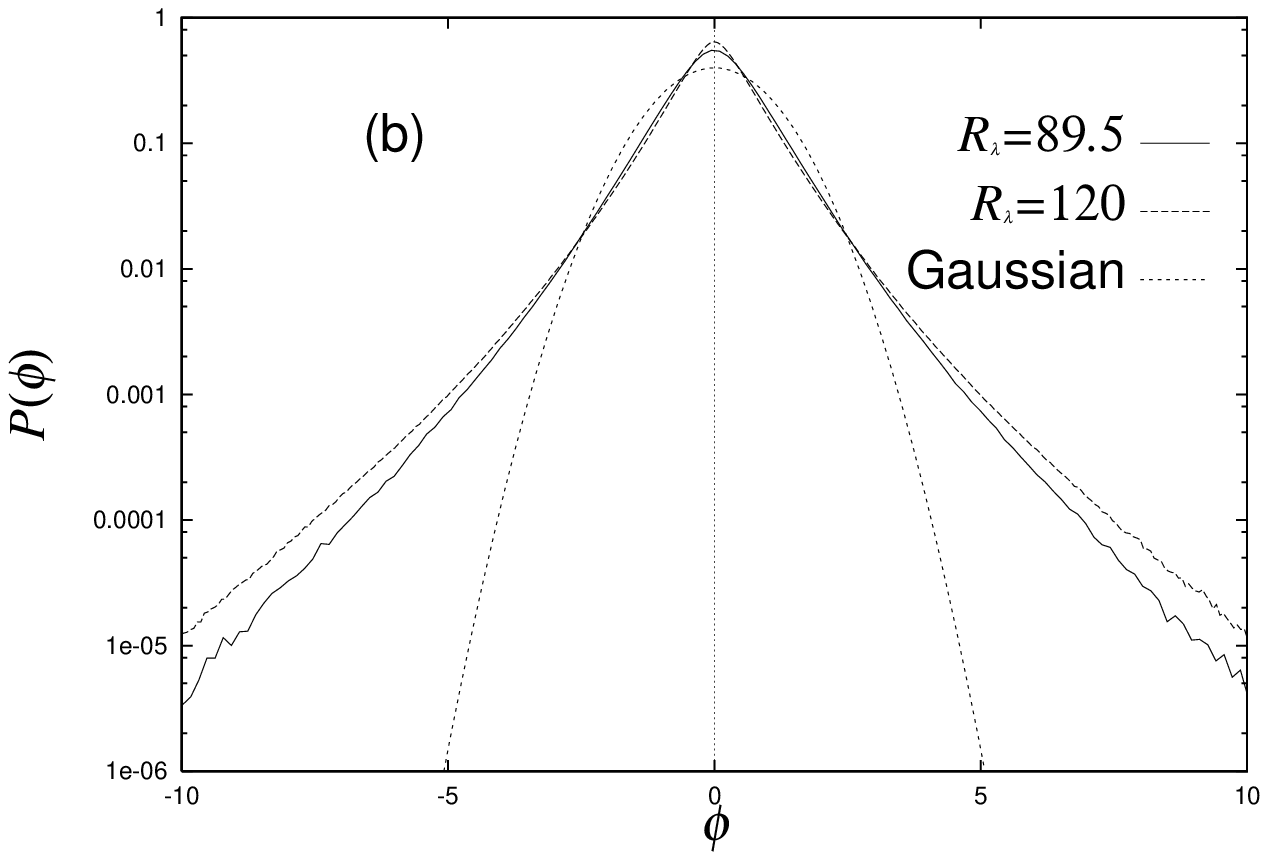,width=75mm}
\caption{The Reynolds number dependence of pdf's of $\theta$ and $\l$ at $r/\eta=20$ in the
inertial region: (a) $P(\theta)$ and (b) $P(\l)$.  The vertical quantities as well as the horizontal ones are
normalized.  The solid line is for $R_{\lambda}=89.5$ in Run 1, and the broken line is for $R_{\lambda}=120$ in Run
3.  The dotted line stands for the Gaussian distribution.  }
\label{fig:2}
\end{figure}

In order to show the Reynolds number dependence of both pdf's, we depicted $P(\t)$ at
$r/\eta=20$ for Run 1 and Run 3 in Fig.\ref{fig:2}(a) and $P(\l)$ in Fig.\ref{fig:2}(b).  The
solid line is for $R_{\lambda}=89.5$ of Run 1, and the broken line is for $R_{\lambda}=120$ of Run
3, together with the dotted line for Gaussian.  The large components of $\t$ and $\l$ become more
frequent as $R_{\lambda}$ increases.

When one compares $P(\l)$ to $P(\t)$ of the same size $r$, for instance by inspection of 
Fig.\ref{fig:1}(a), one notices that $\l$ is more frequent  than $\t$ for large amplitudes and less
frequent for intermediate ones for any $r$.  As $r$ increases, the discrepancy between $\l$ and $\t$
becomes large, although figures associated with the other scales are not shown here.  We suspected
that the difference between both pdf's would be responsible for the deviation of $H$ from the
linearity ansatz, which will be discussed in the following section, but we could not have found a
clear reasoning for it so far.

The Run 2 and 3 have the same Reynolds number $R_{\lambda}=120$, but are simulated on the different 
number of mesh points; the former is on $256^3$ and the latter, on $512^3$.  Thus, the small scale
resolution is less for Run 2 than for  Run 3.  $P(\t)$'s for both runs are almost
the same in the inertial region, but $P(\l)$'s are considerably different from
each other even in the inertial region.  The curve of $H(\Delta)$ for Run 2 is, however, found to
be quite the same in the inertial range as Run 3.  Hence the linearity ansatz (\ref{linear}) may be 
more robust in the inertial region than the pdf's.  

%--------------------------------------------------

\section{Examination of the linear ansatz}

The conditional average $H$ is expressed in the dimensionless form as
\be
   h(\t_r)=H(\Delta_r)\frac{2\kappa \sqrt{S_2(r)}}{J_2(r)}.
\ee 
Note that $h(\t)$ is positively defined for $\t> 0$ because $J_2(r)$ is negative.  Under this 
definition the linear ansatz (\ref{linear}) becomes
\be
   h(\t)=\t.  \label{linear1}
\ee

We investigated $h(\t)$ for various values of $r/\eta$ for every run.
The observed results are summarized.  
\begin{figure}
%\figureheight{6cm}
%\epsfile{file=Fig7plus.epsi,width=75mm}%Thu Oct 30 15:55:17 JST 1997
%\epsfile{file=Fig7plusa.eps,width=75mm}
\psfig{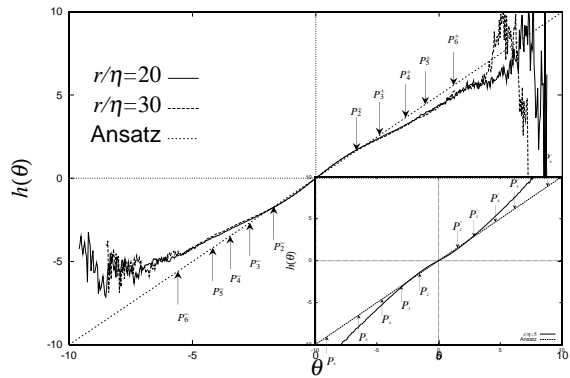}
%\caption{The plot of $\h(\t)$ against $\t$ at $r/\eta=20$ (solid)
\caption{The plot of $h(\t)$ against $\t$ at $r/\eta=20$ (solid)
and 30 (broken) for Run 3.  
The data in the inertial region drop on nearly
the same line.  The dotted line stands for $h(\t)=\t$.  The arrows with $P_n$ have the same
meaning as before.  In the inset $h(\t)$ is plotted for the dissipative
range separation $r/\eta=5$. }
\label{fig:3}
\end{figure}
The linear relation (\ref{linear1}) is approximately valid, but not in detail.  Figure \ref{fig:3} is
the curve of $h(\t)$ against $\t$ for $r/\eta=20$ (the solid line) and $r/\eta=30$ (the broken
line)  for Run 3.  Note that the same type of curve holds for any inertial range separation for other
runs.  The dotted line stands for $h(\t)=\t$. In the inertial  region $h(\t)$ is slightly larger
than $\t$ for small values of $\t$, while for larger values of $\t$ $h(\t)$ is considerably smaller
than the linear relation as seen in Fig.\ref{fig:3}.  Within the present simulations we did not see
the Reynolds number dependence of $h(\t)$, although Ching, L'vov and Procaccia~\cite{rf:clp}
observed by processing the experimental data that the linearity would become more appropriate as the
Reynolds number increases.   

In Fig.\ref{fig:3} the arrows with $P_n$ are inserted for convenience, so that the $n$th order
structure functions are predominantly contributed by the components around $P_n$.  The statistical
reliability is assured  up to $P_6$ or $P_7$ at best.  Note that $h(\t)$ is closer to the linear ansatz about $P_2$,
which is confirmed for any separation in the inertial region.  On the other hand, the linear relation is not good
around $P_n$ with $n \geq 3$.  This suggests that the values of $S_{n}$ with $n \geq 3$ do not rely on the linear
hypothesis in the present simulation. 

The function $h(\t)$ must satisfy the conditions
\begin{subequations}
\be
  \int_0^{\infty} (h(\t)-\t)\t P(\t) d\t=0, \label{condition10}
\ee
and
\be
  \int_0^{\infty} (h(\t)\t -1)P(\t)d\t=0, \label{condition11}
\ee
\end{subequations}
which are derived as follows.  The definition of $H$ yields
\bd
   \int \Delta H(\Delta) P(\Delta) d\Delta= \av{\Delta \Lambda}=-\av{\Sigma}=\frac{J_2}{2\kappa},
\ed
whose dimensionless form reduces to $\int \t h(\t) P(\t) d\t=1$.  On the other hand, $\t$ is normalized so that    
$\int \t^2 P(\t)d\t=1$.  Combining these two equations, we are led to the condition (\ref{condition10}) and
(\ref{condition11}).

The condition (\ref{condition10}) is important in the following sense.  Since $P$ is a functional of $h(\theta)$ and
positive definite,  the function $h(\theta)-\theta$ must change its sign at least  
once if there exists any deviation of $h(\theta)$ from the linear 
ansatz. Function $\theta P(\theta)$ decays rapidly at large $\theta$ 
so that small deviation of $h(\theta)$ from the ansatz at small $\theta$ 
implies large deviation of $h(\theta)$ at large $\theta$. 
Equation (4.3) also can be used as an examination of 
the linear ansatz in that smaller deviation of $h(\theta)-\theta$ from $0$ 
at small $\theta$ where the sample size is usually enough leads to 
smaller deviation at large $\theta$ where the sample size is poor. 
This is consistent with observation in Fig. 3.

In the dissipative range with $r/\eta=5$, $h(\t)$ is given in the inset of Fig.\ref{fig:3}.  The function $h(\t)$
is slightly smaller than $\t$ for small values of $\t$, and larger than one for larger values of
$\t$ in contrast to the inertial range separations.  The crossing of $h(\t)$ with the
linear relation $\t$ occurs around $P_3$ for $r/\eta=5$.  This tendency is observed also in Run 1.

In order to know how much $h(\t)$ is deviated from $\t$ we evaluate a variance
\be
    \sigma=\int^{\infty}_{-\infty} (h(\t)-\t)^2 P(\t) d\t
\ee
for run 3.  At $r/\eta=5$ in the dissipative region $\sigma=0.008$.  As $r$ increases, it decreases and reaches 
zero at $r/\eta=10$, where the linear ansatz is good.   As $r$ increases further in the
inertial range, $\sigma$ grows; $\sigma=0.01$ at $r/\eta=30$.

The observation that $h(\t)$ deviates from the linear relation for $n \geq 3$ in the inertial region 
is understood in the following way.  The ansatz (\ref{linear}) is the projection of the field
$\Lambda(\x, \x')$ on the field $\Delta(\x,\x')$.  The coefficient $f_1(r)$ is obtained by
multiplying eq.(\ref{linear}) by $\Delta(\x,\x')$ and taking the average over the  distribution of
$\Delta(\x,\x')$:
\ba
     f_1(r) S_2(r)&=&\av{\Delta(\x,\x')H(\Delta(\x,\x'))} \no\\
                  &=& \av{\Delta(\x,\x')\Lambda(\x, \x')}=\frac{J_2(r)}{2 \kappa},
\ea
from which we have $f_1(r)=J_2(r)/(2 \kappa S_2(r))$.  The calculation of $J_2(r)$ is the same as that of the 
projection coefficient.   Since $J_2(r)$ is directly related to $S_2(r)$, it is naturally expected that the linear
ansatz including the definition of $f_1(r)$ is well satisfied at $n=2$, i.e. around $P_2$.    Let us go to
larger values of $n$.  In the calculation of $J_{n}(r)$ the field $\Lambda(\x,\x')$ must be
projected on the space where the quantity
$\Delta(\x,\x')^{n-1}$ is dominant.  Such a space is considerably localized, so that the projection 
coefficient $f_1(r)$ is not suitable.  $f_1(r)$ must be replaced by
\be
     \tilde{f}_1(r)=\frac{\tilde{J}_2(r)}{2 \kappa \tilde{S}_2(r)},
\ee
where $\tilde{J}_2(r)$ and $\tilde{S}_2(r)$ are computed over such an enhanced space.  
Since the equation for the scalar field is linear, one may expect that the ratio of $J_2(r)$ to $S_2(r)$ is
constant everywhere.  This would be correct if the scalar field did not move in space and the
cascade occurred locally in space.  While the strong fluctuations of scale $r$ cascade down to the
dissipation scale where the field $\Lambda$ is mainly contributed, they will diffuse to the less
intense space through the convection term due to the velocity field and the viscous term 
due to the molecular viscosity.

Although $\tilde{J}_2(r)$ is larger than $J_2(r)$, a ratio of $\tilde{J}_2(r)$ to $J_2(r)$ is less than 
that of $\tilde{S}_2(r)$ to $S_2(r)$.  Hence $\tilde{f}_1(r)$ is expected to be less than $f_1(r)$, 
implying that $h(\t)$ is lower than $\t$.  On the other hand, weak fluctuations will be affected 
in the opposite way, which may be responsible for the positive deviation from the linearity.

\begin{figure}
%\figureheight{6cm}
%\epsfile{file=tghRUN3re20n2.eps,width=75mm}%NOT YET
\psfig{file=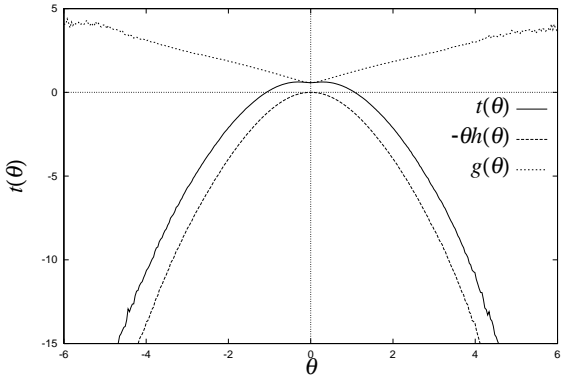,width=75mm}%NOT YET
\caption{%
The conditional average of the spatial transfer of the energy flow, i.e.
$t(\t)=g(\t)-\t h(\t)$ for $r/\eta=10$ in Run 3.  The solid line stands for $t(\t)$, the broken one does for $-\t
h(\t)$, and the dotted one does for $g(\t)$.  A positive value of $t(\t)$ corresponds to the inflow and a negative
one does to the outflow. }
\label{fig:4}
\end{figure}

The above interpretation can be supported more directly by investigating the conditional average of 
the following identity with a fixed value of $\Delta(\x,\x+\r)$:
\ba
  & &  \Delta(\x,\x+\r) \nabla_{\x}^2 \Delta(\x,\x+\r)= \no\\
  & &    \hspace{2ex} \nabla_{\x} \cdot (\Delta(\x,\x+\r) \nabla_{\x} \Delta(\x,\x+\r))
          -|\nabla_{\x} \Delta(\x,\x+\r)|^2. \no
\ea
Using the notation introduced previously, we have
\ba
  & &  \hspace{-1ex}\av{\nabla_{\x} \cdot (\Delta(\x,\x+\r) \nabla_{\x} \Delta(\x,\x+\r))|\Delta(\x,\x+\r)} \no\\
  & &  \hspace{-2ex} =\Delta(\x,\x+\r)H(\Delta(\x,\x+\r))+G(\Delta(\x,\x+\r)), \label{transfer}
\ea
where the left hand side is the conditional average of the spatial transfer rate with fixed 
$\Delta(\x,\x')$.  Since the two conditional averages on the right hand side are numerically
evaluated, the dimensionless conditional transfer rate $t(\t) \, (\equiv g(\t)-\t h(\t))$ is calculated as
given in Fig.\ref{fig:4}.  The region with positive  values of $t(\t)$ is characterized by the
inflow, while that with negative values is by the outflow.  About $P_2$, $t(\t) \sim 0$ as expected. 
For large values of $\Delta(\x,\x')$ the outflow is dominant.  Hence $\tilde{f}_1(r) < f_1(r)$ is
anticipated in such a region.  For small values of
$\Delta(\x,\x')$ the inflow is dominant, so that the opposite tendency is expected.  

\section{Examination of the second conditional average}

We studied $G(\Delta_r)$, the conditional average of $\Sigma(\x,\x+\r)$ with
$\Delta(\x,\x+\r)$ fixed.  For simplicity we express $G$ in the dimensionless
form
\be
    g(\t_r)=-\frac{2\kappa}{J_2(r)}G(\Delta_r). 
\ee
The analytic expression $g(\t)$, if possible, is very useful to discuss the pdf $P(\t)$, since it is directly related
%to the pdf as shown later. %Thu Oct 30 15:55:51 JST 1997
to the pdf as shown in the following section.
 Hence we try to investigate the funtional form of $g(\t)$ in detail.  

\begin{figure}
%\figureheight{12cm}
%\epsfile{file=Fig9rev.eps,width=75mm}
\psfig{file=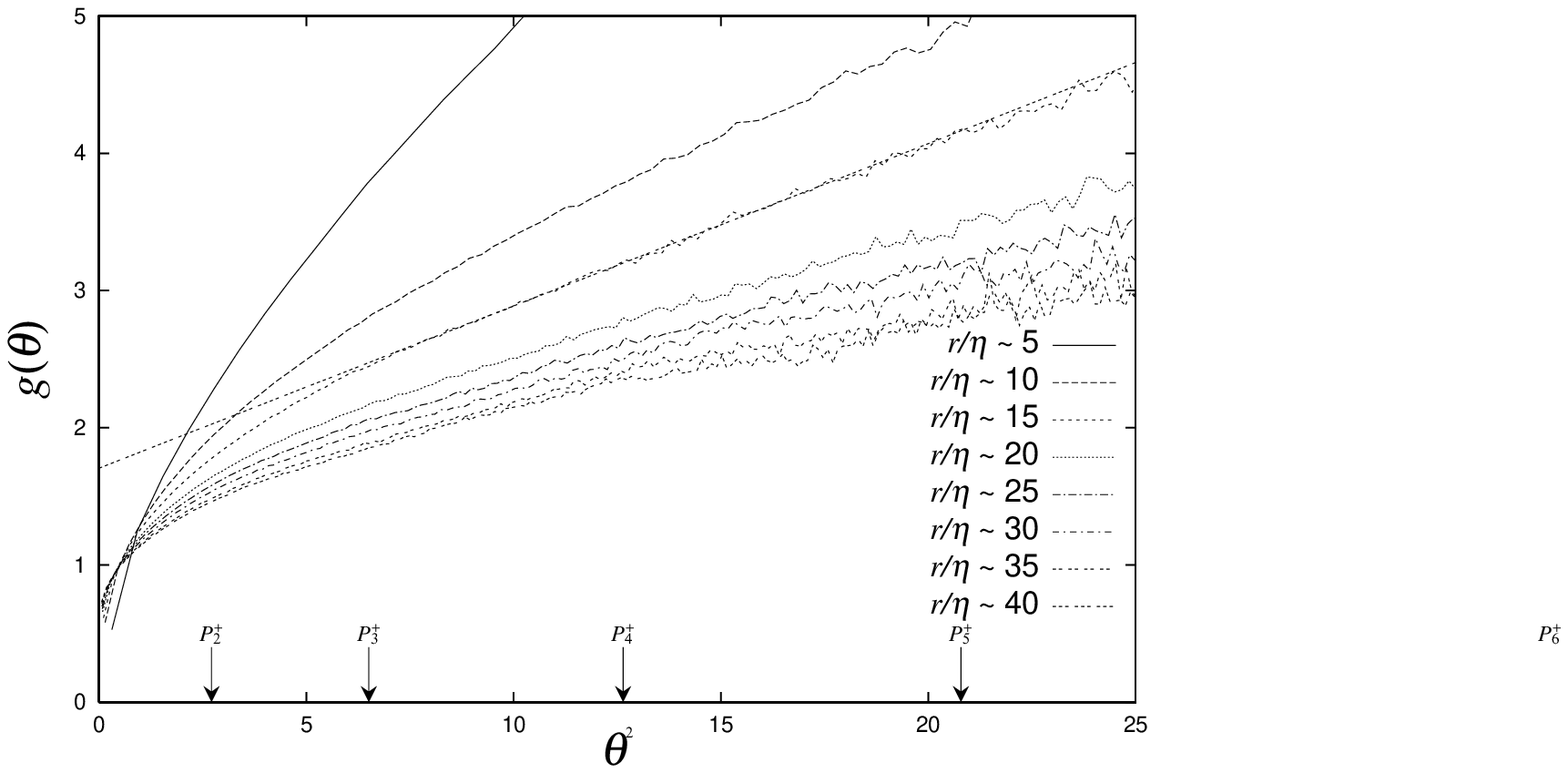,width=75mm}
%
%\epsfile{file=Fig12.eps,width=75mm}
\caption{ The plot of $g(\t)$ for various $r$ for Run 3.  It is plotted against $\t^2$.  The
linearity is quite good for $r/\eta=15$ as seen from the dotted straight line $1.7+0.12\t^2$ drawn on
the curve for $r/\eta=15$.  }
\label{fig:5}
\end{figure}

\begin{figure}
%\epsfile{file=Fig11.eps,width=75mm}
\psfig{file=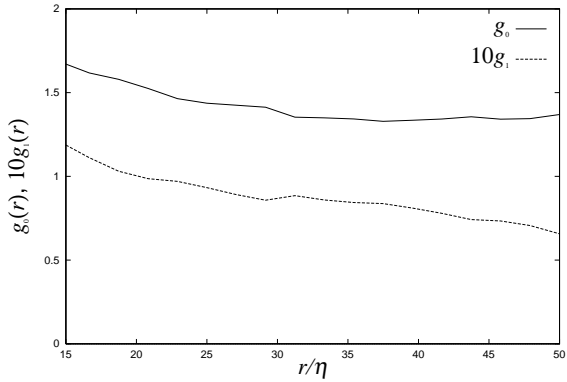,width=75mm}
%\figureheight{6cm}
\caption{%
The plot of $g_0(r)$ and $g_1(r)$, which are estimated from
 Fig.\ref{fig:5} in the range $7 \leq 
\t^2 \leq 13$, as a function of $r/\eta$ in the inertial region for Run
 3.  The solid line is $g_0(r)$,
while the broken line is $10 g_1(r)$. }
\label{fig:6}
\end{figure}

First we plot $g(\t)$ against $t^2$.  The reason for it is that there must be a certain conservation 
relation between the scalr dispersion $\t^2$ and the scalr diffution rate $g(\t)$ for large values of $\t$. 
Figure \ref{fig:5} is such a plot for $|\t|<5$.  According to it, $g(\t)$ may be 
fitted by a quadratic form
\be
   g(\t)=g_0(r)+g_1(r) \t^2  \label{fitg}
\ee
for large values of $\t$.  For
$r/\eta=5$ in the dissipative region the curve is convex, so that (\ref{fitg}) is not valid; $g(\t)
\sim \t^{\alpha}$ with $\alpha <2$.  In the inertial region $r/\eta=15$ and 20 we can see the
considerable range fitted by (\ref{fitg});  the dotted straight line drawn on the curve for 
$r/\eta=15$ is expressed by $1.7+0.12 \t^2$.  The linearity is remarkable in the range between
$P_3$ and $P_5$ (which are calculated for $r/\eta=15$), where the linear ansatz (\ref{linear}) fails,
as seen in Fig.\ref{fig:3}.  Such a fitting range becomes short as $r$ increases.  By fitting the
formula (\ref{fitg}) to the curve of $g(\t)$ of Fig.\ref{fig:5} in the range $7 \leq \t^2 \leq 13$, we
estimated $g_0(r)$ and $g_1(r)$ as a function of $r$; they are depicted in Fig.\ref{fig:6}; the solid
line is for $g_0$ and the broken one is for $10 g_1$.  Both are
slightly decreasing with $r$.  Note that $g_1(r)$ is one order of magnitude less than $g_0(r)$.
\begin{figure}
%\epsfile{file=Fig12rev.eps,width=75mm}
\psfig{file=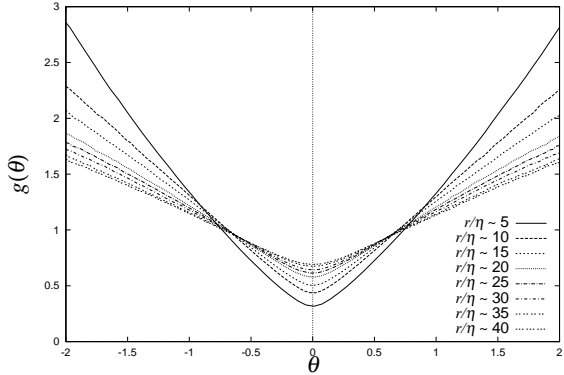,width=75mm}
%\figureheight{6cm}
\caption{ The plot of $g(\t)$ for various $r$ for Run 3. It is plotted against $\t$ in the range of small values of
$\t$. }
\label{fig:5b}
\end{figure}

In the region with small values of $\t$ $g(\t)$ deviates from a fitting formula (\ref{fitg}).  To
clarify the point, we depicted $g(\t)$ against $\t$ in  Fig.\ref{fig:5b} in the range $|\t| \leq 2$. 
This figure indicates that the curve is described by a quadratic form in the vicinity of $\t \sim 0$
and then, by the linear form for larger values of $\t$ around 1.  The linear region may be a
transition region between the two different quadratic forms.

Before concluding this section we want to add that the function of
$g(\t)$ must satisfy the relation
\be
   \int (g(\t)-1)P(\t)d\t = \int (g(\t)-\t^2)P(\t)d\t=0, \label{condition2}
\ee
which is derived in a similar way to (\ref{condition10}).  From the definition of $G$, its dimensionless form satisfies
\bd
   \int g(\t) P(\t) d\t=1.
\ed
Combining it with $\int P(\t)d\t=\int \t^2P(\t)d\t=1$, we have the condition (\ref{condition2}).  It is guaranteed in
Fig.5 and 7.

\section{Derivation of the PDF of $\t$ in terms of $h(\t)$ and $g(\t)$}

Ching and Kraichnan~\cite{rf:ck} argued that $P(\t)$ in the homogeneous system is expressed 
in terms of $h(\t)$ and $g(\t)$ through
\be
    P(\t)g(\t)=C \exp \left( -\int^{\t}_0 du \, q(u) \right), \label{P}
\ee
where $C$ is a normalizing constant and 
\be
    q(u)= \frac{h(u)}{g(u)}.
\ee

Since the derivation of the relation is straightforward and short, we repeat their derivation for convenience.
The extension of (\ref{transfer}) to any $n$ yields
\ba
  & & \av{\nabla_{\x} \cdot (\Delta(\x,\x+\r)^{n-1} \nabla_{\x} \Delta(\x,\x+\r))|\Delta(\x,\x+\r)} \no\\
  & & \hspace{5ex} =\Delta^{n-1}(\x,\x+\r)H(\Delta(\x,\x+\r)) \no\\
  & & \hspace{5ex} +(n-1)\Delta^{n-2}(\x,\x+\r)G(\Delta(\x,\x+\r)) 
\ea
If the system is homogeneous in space, the average on the left hand side will be replaced by the spatial average, 
so that the left hand side vanishes.  Then in the dimensionless form this equation becomes
\ba
   \int d \t \t^{n-1} h(\t) P(\t)
        &=&(n-1) \int d \t \t^{n-2} g(\t) P(\t) \no\\
        &=& -\int d \t \t^{n-1} \frac{d}{d\t}g(\t) P(\t).
\ea
The above relation must hold for any $n$, so that
\be
   h(\t) P(\t)=-\frac{d}{d\t}g(\t) P(\t),
\ee
whose solution is (\ref{P}).  Hence the relation is expected to be exact for a spatially homogeneous
system.  Note that the similar expression for the pdf was proposed already by Yakhot~\cite{rf:ya} in   
B\'{e}nard convection.

\begin{figure}
%\epsfile{file=ZsmRUN1.eps,width=75mm}
%
%\epsfile{file=gPRUN3.eps,width=75mm}
\psfig{file=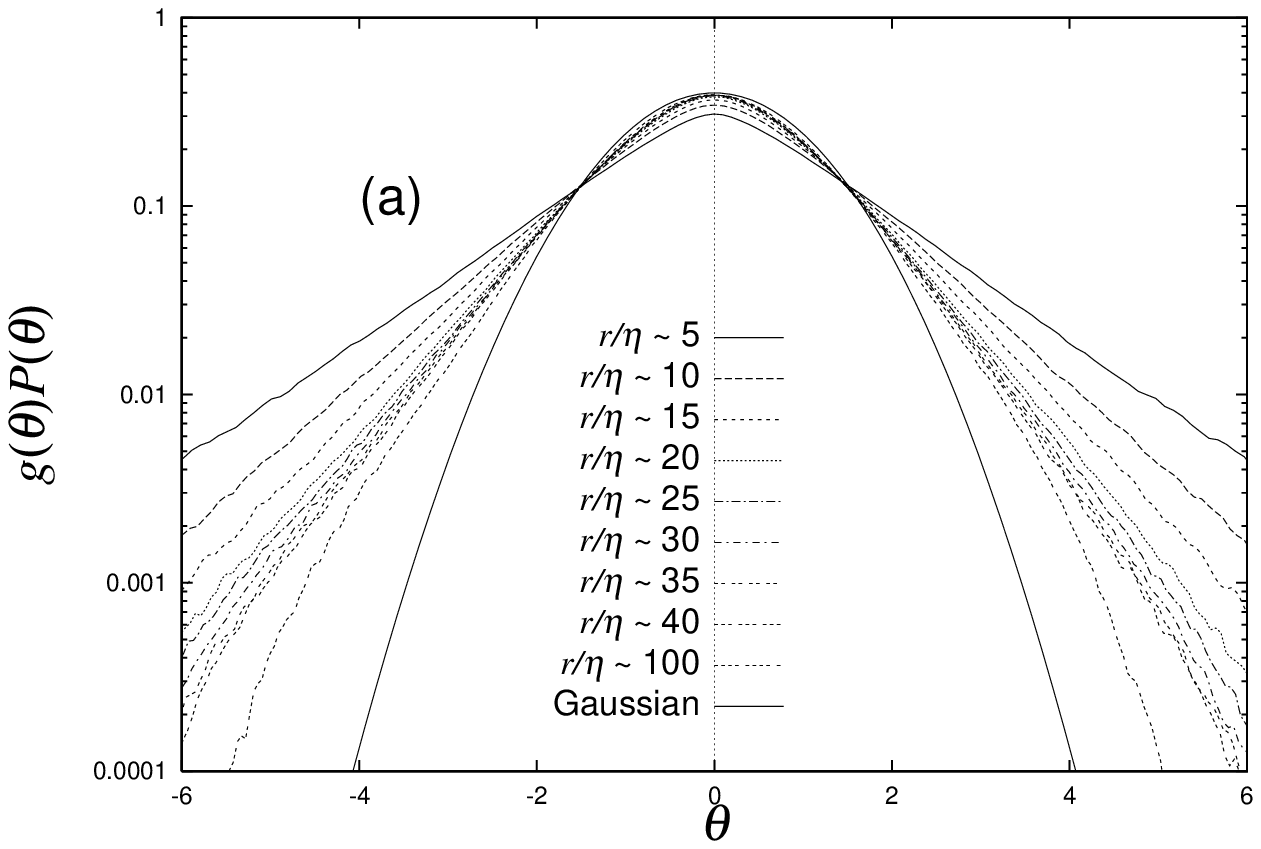,width=75mm}

\psfig{file=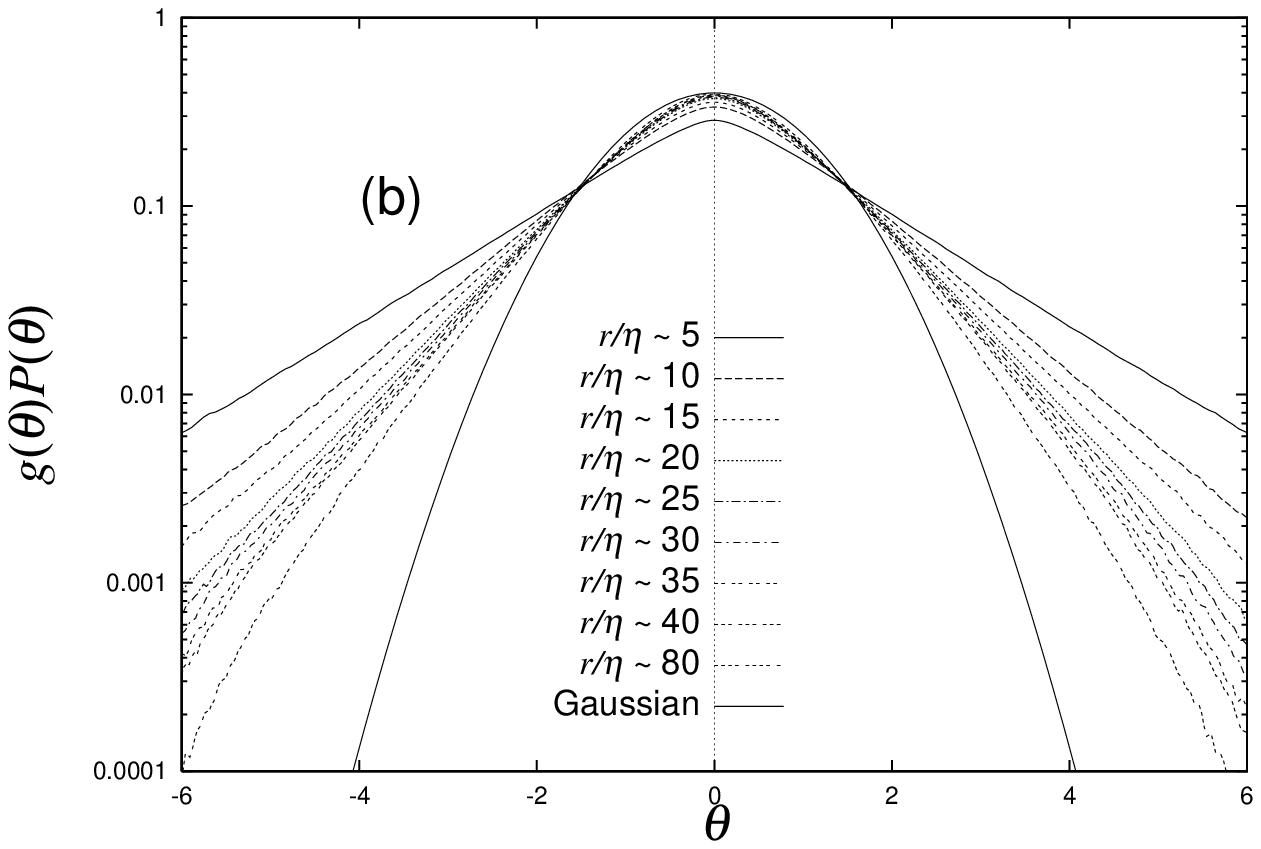,width=75mm}
%\figureheight{12cm}
\caption{The product of $P(\t)$ and $g(\t)$ for various values of $r/\eta$.  The
Gaussian pdf is also inserted.  (a) Run 1 and (b) Run 3. 
}
\label{fig:7}
\end{figure}

First, we present the observed result of $P(\t)g(\t)$ for Run 1 in Fig.\ref{fig:7}(a) 
and for Run 3 in Fig.\ref{fig:7}(b) for various values of $r$ ranging from $r/\eta=5$ in the 
dissipative region to $r/\eta=100$ in the energy containing region.  The Gaussian pdf is
also added.   (1) The product form $P(\t)g(\t)$ looks simpler than $P(\t)$ itself in
Fig.\ref{fig:1}.  The fitting formula for $P(\t)g(\t)$ will be given later.  
(2) All curves pass through an almost common point.  According to Fig.\ref{fig:7}(a) and (b) the point
is read as $(\t=1.5,\, P(\t)g(\t)=0.13)$.   

When we magnified the crossing region of the curves, however, we noticed that the crossing does not occur at a single
point.  All the curves for the inertial region ranging from $r/\eta=15$ to 50 (Run 3) cross at $\t=1.31$.  On the
other hand, the curves for the energy containing region and the dissipation region meet those for the inertial region 
at slightly larger values of $\t$, i.e. about $(1.4 \sim 1.5)$.

\begin{figure}
%\figureheight{6cm}
%\epsfile{file=qRUN3.eps,width=75mm}
\psfig{file=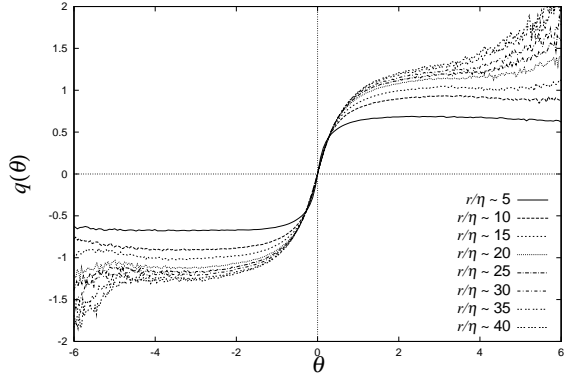,width=75mm}
\caption{
The function $q(\t)$, which is numerically evaluated with the
aid of the observed $h(\t)$ and $g(\t)$ for Run 3, is plotted against
$\t$ for various values of $r/\eta$.
} 
\label{fig:8}
\end{figure}

Second, we pay attention to the function $q(\t)=h(\t)/g(\t)$.   Substituting the
measured values of $h(\t)$ and $g(\t)$ into $q(\t)$, we numerically obtained $q(\t)$ as a function of
$\t$; the function $q(\t)$ for various  values of $r/\eta$ for Run 3 is given in Fig.\ref{fig:8}.  It
increases linearly from the origin.  The slope decreases with increasing $r$, but the
decrease is not appreciable in the inertial region.  For the separation in the dissipative region,
$q(\t)$ saturates and then, decreases very slowly.  As $r$ increases, it does not saturate
enough.  The curves lose the symmetry a little when $|\t|$ goes beyond 4.  When $r$ increases, the
considerable fluctuations are observed in the region with large values of $\t$.  
Since $q(\t)$ looks simpler than $h(\t)$ and $g(\t)$ themselves, the fitting formula for $q(\t)$ is
more suitable, as developed later.  

\begin{figure}[tbh]
%\epsfile{file=gPqRUN3reta05.eps,width=75mm}
%
%\epsfile{file=gPqRUN3reta20.eps,width=75mm}
\psfig{file=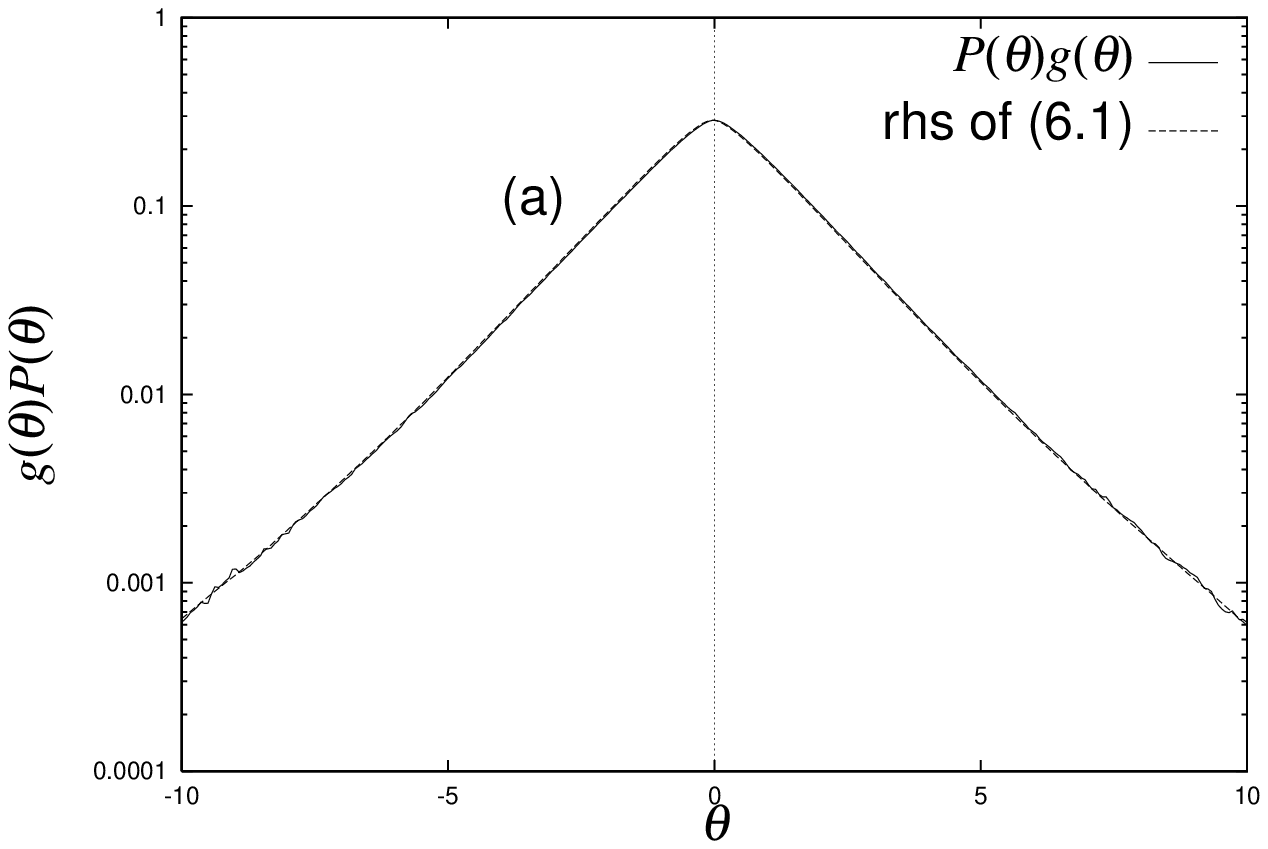,width=75mm}

\psfig{file=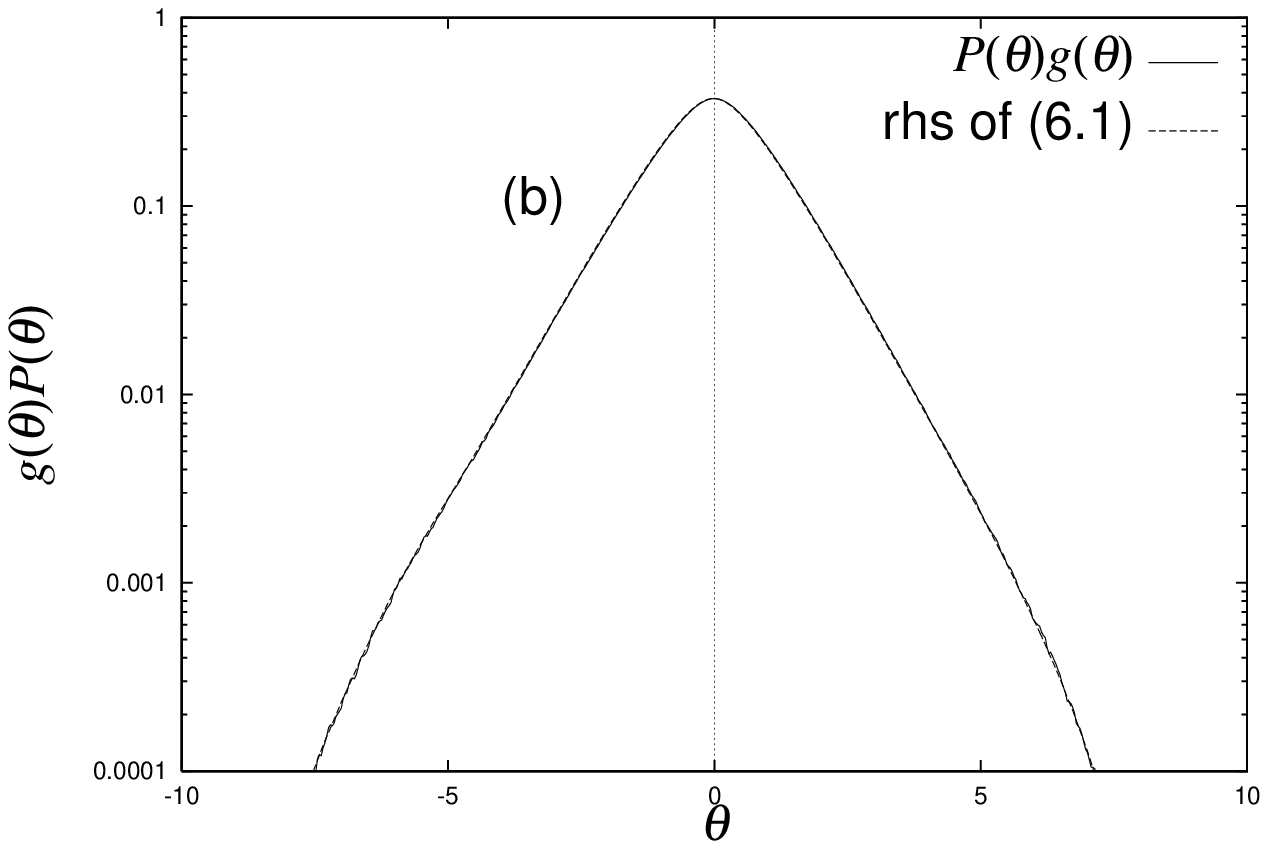,width=75mm}
%\figureheight{12cm}
\caption{%
The comparison of the numerically integrated right hand side
of eq.(6.1) with the observed pdf multiplied by $g(\t)$ for
(a)$r/\eta=5$ and (b) $r/\eta=20$ for Run 3. 
}
\label{fig:9}
\end{figure}

Since the simulated system is considered to be homogeneous in space, the relation (\ref{P}) is expected to hold.
Substituting the calculated $q(u)$ into the right hand side of (\ref{P}) and carrying out the numerical 
integration over $u$, we evaluated the right hand side of (\ref{P}).  Then the result was compared
with the observed $P(\t)$ multiplied by $g(\t)$.  The comparison was shown for $r/\eta=5$ and $r/\eta=20$
of Run 3 in Fig.\ref{fig:9}(a) and \ref{fig:9}(b).  The agreement is perfect.

Before going to discuss the analytical form for $q(\t)$ and $g(\t)P(\t)$, we would like to consider 
which part of fluctuations contributes to $q(\t)$, and $g(\t)P(\t)$ consequently.  Let us start
with small values of $\t$, where $h(\t)=\t$ and $g(\t)=\mbox{const}=c_0$.  Hence, $q(u)=u/c_0$,
leading to the Gaussian distribution according to (\ref{P}) and the curvature is inversely
proportional to $c_0$.  Since $c_0$ decreases as $r$ decreases, as seen from Fig.7, the
curvature around $\t=0$ is a decreasing function of $r$, in agreement with Fig.\ref{fig:7}.  

The deviation from Gaussian begins to occur as $\t$ grows.  As $\t$ increases, $h(\t)$ is still close to $\t$, but
$g(\t)$ is not constant.
%Figure \ref{fig:5}%Thu Oct 30 16:00:27 JST 1997
Figure \ref{fig:5b}
suggests that $g(\t)$ is approximated by 
$g(\t)=c_0+c_1 \t$ in that region, where $c_0$ increases and $c_1$ decreases with increasing $r$. 
Then 
\be
   g(\t)P(\t)=C \exp \left( -\frac{\t}{c_1}+\frac{c_0}{c^2_1} \log \frac{\t+c_0/c_1}{c_0/c_1} \right).
\ee
The existence of $c_1$ makes the pdf deviate from Gaussian.  If the first term is dominant, the pdf
is of the exponential type: the exponent $1/c_1$ increases with $r$, in consistent with the 
observation of Fig.\ref{fig:7}. 

\begin{figure}
%\epsfile{file=bfreBRUN3.eps,width=75mm}
\psfig{file=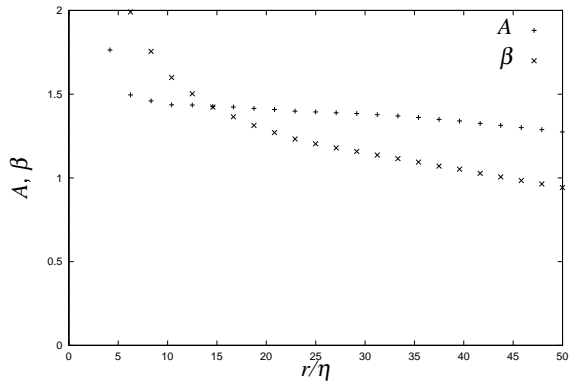,width=75mm}
%\figureheight{6cm}
\caption{%
The fitting parameters $A$ and $\beta$ as a function of $r$ for Run 3; the crosses stand for $\beta$,
while the pluses do for $A$. }
\label{fig:10}
\end{figure}

To conclude this section, we would like to propose an analytic expression of $q(\t)$.  The analytical 
expression, if possible, is helpful for constructing a turbulence model.  The combined function
$q(\t)$ is simpler than $h(\t)$ and $g(\t)$ themselves.  It may be approximately parametrized as 
\be
   q(\t)=\frac{A}{\beta} \tanh(\beta \t), \label{q}
\ee
where $A$ and $\beta$ depend on $r$.  The most optimal fitting values for them are depicted in Fig.
\ref{fig:10}; the fitting was done in the symmetric region $|\t| \leq 4$.  The crosses stand for
$\beta$, while the pluses do for $A$; $A$ and $\beta$ are decreasing with $r$, while $A/\beta$ is
increasing.  Although the best fit for the entire region of $\t$ is given in that figure, the local
fitting is also possible.  (For instance, the separate  regions $\t \leq 1$ and $\t \geq 1$ would be
fitted more appropriately.)  

Substituting (\ref{q}) into (\ref{P}) yields
\be
    P(\t)g(\t)=C (\cosh \beta \t)^{-A/\beta^2}, \label{pg}
\ee
where $C=\beta/B(A/2\beta^2,1/2)$, with $B$ being a beta function.  For small $\t$, $P(\t)g(\t) 
\sim \e^{-A \t^2/2}$, i.e. Gaussian.  The curvature decreases as $A$ when $r$ increases, in
agreement with Fig.\ref{fig:7}.   For large $\t$, it becomes
$\e^{-A\t/\beta}$, i.e. the pdf is of the exponential type, whose coefficient is $A/\beta$, which is 
an increasing function of $r$, also consistent with Fig.\ref{fig:7}. 

\begin{figure}
%\epsfile{file=pgqRUN3reta10.eps,width=75mm}
\psfig{file=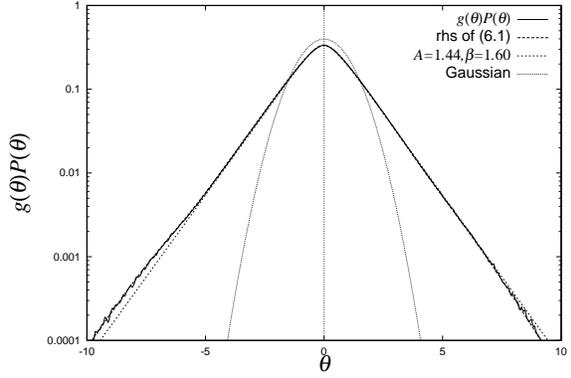,width=75mm}
%\figureheight{6cm}
\caption{ The comparison of (\ref{pg}) to the observed function of $g(\t)P(\t)$ at $r/\eta=10$.  The 
solid line stands for the observed value of $g(\t)P(\t)$ and the broken line does for the right hand
side of (\ref{P}) obtained after integrating $q(u)$ over $u$.  The dotted line is given by (\ref{pg})
with $A=1.44$ and $\beta=1.60$.}
\label{fig:11}
\end{figure}

In order to show how the fitting works, we would like to give $g(\t) P(\t)$ for $r/\eta=10$ as an example  in
Fig.\ref{fig:11}, where $A=1.44$ and $\beta=1.60$ for $r/\eta=10$ are employed from Fig.\ref{fig:10}. 
The solid line stands for the observed value of $g(\t)P(\t)$ 
%in Fig.8 and the broken line does for the right%Thu Oct 30 15:57:38 JST 1997
in Fig.8(b) and the broken line does for the right
hand side of (\ref{P}) obtained after integrating $q(u)$ over $u$.  The dotted line is given
by (\ref{pg}) with selected $A$ and $\beta$.  The agreement is very satisfactory.  

\section{Discussion}

When the equation (\ref{dj}) is expressed through the pdf,
it takes the form
\ba
   & &  -r^{1-d} \frac{\partial}{\partial r} \left( a(r) r^{d-1} \frac{\partial}{\partial r} 
            P(\Delta,r) \right) \no\\
   &=&  -\kappa \frac{\partial}{\partial \Delta} (P(\Delta,r)H(\Delta,r)) \no\\
   &=&  -\kappa \frac{\partial^2}{\partial \Delta^2 } (P(\Delta,r)G(\Delta,r)). 
\ea
If the velocity field is not $\delta$-correlated in time, the inertial term will be more
complicated.   In any case, however, $P(\Delta), H(\Delta)$ and $G(\Delta)$ must satisfy  similar 
equations.   In this sense the information on $H$ and $G$ extracted from a study of
the simulation is useful to construct a model for turbulence.  For this purpose we summarize the
useful points.

\begin{enumerate}
\item The combined form $q(\t)$ is a more appropriate function for modeling, rather than $h(\t)$ and 
$g(\t)$ themselves.
\item Another combined form $P(\t) g(\t)$ is more appropriate than the pdf $P(\t)$ itself.  
$P(\t)g(\t)$ is the weighted conditional average of the scalar diffuison rate $g(\t)$.  
\item The curves $P(\t) g(\t)$ with varying $r$ drop on an almost common point.
\end{enumerate}

Let us consider the meaning of $g(\t)$.  Take a specific scale $r$.  Even if $\t_r$ is fixed, 
there are many structures, so that the scalar diffuison rate $|\nabla_{\x} \t(\x,\x+\r)|^2$ vary from
structure to structure.  After taking the average over structures,
we have the conditionally averaged dissipation rate $g(\t_r)$.

The $g_r(\t_r)P_r(\t_r)$ is the average %
%dissipation%Thu Oct 30 16:45:57 JST 1997
diffusion
rate at scale $r$ weighted by the frequency
$P_r(\t_r)$ of $\t_r$.   Here we insert the subscript $r$ for the functions $g$ and $P$ to show their
scale dependence explicitly.  The present simulation indicates that $\tilde{g}_r(\t_r)
(\equiv g_r(\t_r)P_r(\t_r))$ takes the same value $0.13$ at $\t_r=1.5$ irrespective of $r$.  For $\t_r
\leq 1.5$  $\tilde{g}_r(\t_r)$ increases with increasing $r$, while it decreases for $\t_r \geq
1.5$.  This is surprising, we think.  Why?  To show it, we consider two specific scales $r_1$ and
$r_2$ such that $r_1<r_2$.  Then we can have a situation that $\tilde{g}_{r_1}(\t) \leq
\tilde{g}_{r_2}(\t)$ for $\t \leq \t_c$, where $\t_c$ is a certain value, while $\tilde{g}_{r_1}(\t)
\geq \tilde{g}_{r_2}(\t)$ for $\t \geq \t_c$.  However, the present DNS suggests that this
inequality occurs at the same $\t_c=1.3$ for any inertial separation $r$, ranging from the dissipative to energy
containing scale.

Finally we speculate the pdf in turbulence for large Reynolds numbers based on the results, although the present
DNS decaying turbulence has only a limited interval of the inertial range.  In this paper we observed the region where
$g(\t) \propto \t^2$ for large values of $\t$.  If such a term is dominant,
and $h(\t)$ is $\t$, the function $q(\t)$ is proportional to $\t^{-1}$, yielding the power
law distribution as suggested by Sinai and Yakhot,~\cite{rf:sy} and Ching.~\cite{rf:c}  Figure
\ref{fig:5} indicates that the dominance of the square term over the constant term
in $g(\t)$ becomes more enhanced as $r$ decreases, so that the pdf of the smallest scale is expected
to become more power-law like.  However, in this simulation the pdf for the smallest scale is 
still of exponential type.  The reason for it is that $h(\t) \sim \t$, but $g(\t)$ is weaker than
$\t^2$, as seen in Fig.\ref{fig:5}, so that $q(\t)$ at $r/\eta=5$ decreases slightly for large
$\t$ (not as $1/\t$), as seen in Fig.\ref{fig:8}.  In order to know the behavior of the pdf for
large amplitude, therefore, the information of $q(\t)$ is more indispensable than that of
individual conditional average $h(\t)$ and $g(\t)$.

\end{document}